\documentclass[aps,twocolumn,preprintnumbers,amsmath,amssymb,superscriptaddress,floatfix,nofootinbib]{revtex4}

\pdfoutput=1

\usepackage{lipsum}
\usepackage{graphicx,color}
\usepackage{epsfig}
\usepackage{epstopdf}
\usepackage{hyperref}
\usepackage{amsmath}
\usepackage{amsfonts}
\usepackage{amssymb}

\usepackage{float}

\begin{document}

\title{Considerations on the Schmid theorem for triangle singularities}

\author{V.~R.~Debastiani}
\email{vinicius.rodrigues@ific.uv.es}
\affiliation{Departamento de F\'{i}sica Te\'{o}rica and IFIC, Centro Mixto Universidad de Valencia - CSIC,
Institutos de Investigaci\'{o}n de Paterna, Aptdo. 22085, 46071 Valencia, Spain.}

\author{S.~Sakai}
\email{shsakai@itp.ac.cn}
\affiliation{Institute of Theoretical Physics, CAS,
Zhong Guan Cun East Street 55
100190 Beijing, China}
\affiliation{Departamento de F\'{i}sica Te\'{o}rica and IFIC, Centro Mixto Universidad de Valencia - CSIC,
Institutos de Investigaci\'{o}n de Paterna, Aptdo. 22085, 46071 Valencia, Spain.}

\author{E.~Oset}
\email{eulogio.oset@ific.uv.es}
\affiliation{Departamento de F\'{i}sica Te\'{o}rica and IFIC, Centro Mixto Universidad de Valencia - CSIC,
Institutos de Investigaci\'{o}n de Paterna, Aptdo. 22085, 46071 Valencia, Spain.}

\date{\today}

\begin{abstract}
We investigate the Schmid theorem, which states that if one has a tree level mechanism with a particle decaying to two particles and one of them decaying posteriorly to two other particles, the possible triangle singularity developed by the mechanism of elastic rescattering of two of the three decay particles does not change the cross section provided by the tree level. We investigate the process in terms of the width of the unstable particle produced in the first decay and determine the limits of validity and violation of the theorem. One of the conclusions is that the theorem holds in the strict limit of zero width of that resonance, in which case the strength of the triangle diagram becomes negligible compared to the tree level. Another conclusion, on the practical side, is that for realistic values of the width, the triangle singularity can provide a strength comparable or even bigger than the tree level, which indicates that invoking the Schmid theorem to neglect the triangle diagram stemming from elastic rescattering of the tree level should not be done.
Even then, we observe that the realistic case keeps some memory of the Schmid theorem, which is visible in a peculiar interference pattern with the tree level.
\end{abstract}

\keywords{Triangle Singularities, Schmid Theorem, Resonances, Rescattering, Kinematics.}

\maketitle

\section{Introduction}

Let us look at a process proceeding at tree level, depicted in Fig.~\ref{fig:1}, in which particle $A$ decays into $R$ and $1$, and $R$ further decays into particles $2$ and $3$.
Triangle singularities stem from the related reaction mechanism, which is depicted in Fig.~\ref{fig:2}, where the particle $A$ decays into $R$ and $1$, posteriorly $R$ decays into $2$ and $3$ and later $1$ and $2$ fuse to give a new particle or simply rescatter to give the same state (or another one if there are inelasticities). Diagrammatically the process is depicted in Fig.~\ref{fig:2} where there is a loop containing particles $1$, $2$ and $R$ as intermediate states. This loop function can lead to some singularities (for the limit of zero width of the $R$ particle) when all the three intermediate particles are placed on shell, particle $1$ and $3$ are antiparallel and furthermore a condition is fulfilled which in classical terms can be stated as corresponding to the decay of $A$ at rest into $R$ and $1$, assuming $R$ moves forward and $1$ backward, then $R$ decays into $2$ and $3$, and letting $3$ go forward, $2$ eventually goes backward, which means in the same direction of $1$. If $2$ moves faster than $1$ it can catch it and either rescatter or fuse. These conditions are the essence of Coleman-Norton theorem \cite{Coleman}.
Early studies of triangle singularities were done in Ref.~\cite{Karplus} but the systematic study was done by Landau \cite{Landau}.

Work followed in Refs.~\cite{Peierls,Aitchison,Chang,Bronzan} and some peaks observed in nuclear reactions \cite{Booth} were suggested as indicative of a triangle singularity \cite{Dakhno}.
A thorough discussion of this early work was done by Schmid in a clarifying article \cite{Schmid}. There a surprise appeared, known nowadays as Schmid theorem that states that if the rescattering of particles $1$ and $2$ occurs, going to the same state $1+2$, the triangle singularity does not lead to any observable effect in magnitudes like cross sections or differential widths. It is simply reabsorbed by the $S$-wave of the tree level amplitude (the same mechanism without rescattering, see Fig.~\ref{fig:1}) modifying only the phase of this partial amplitude. In angle-integrated cross sections the effect of the triangle singularity disappears.

Some debate originated on the limits to the Schmid theorem and its range of validity \cite{Anisovich,Watson,Aitchison2,Adam}. In Ref.~\cite{Anisovich}, for example, it was shown that if the scattering amplitude of $1+2 \to 1+2$ contains inelasticities then the Schmid theorem does not hold.

Recently there has been a renewed interest in triangle singularities because the present wealth of experimental work offers multiple possibilities to study such mechanisms.
A topic that stimulated the present interest on the subject was the one of isospin violation in the $\eta(1405)$ decay into $\pi^0 f_0(980)$ versus the isospin allowed decay into $\pi^0 a_0(980)$ \cite{Wu,AcetiWu,WuWu,Achasov1,Achasov2}.
One interesting work was done in Ref.~\cite{Zhao}, with many suggestions of places and reactions where triangle singularities could be found.
One of the most striking examples of this rebirth is the work of Refs.~\cite{Mikhasenko,Aceti} where a peak observed by the COMPASS collaboration \cite{COMPASS}, and branded as a new resonance $a_1(1420)$, was shown to be actually produced by a triangle singularity. The renewed interest in the issue was also spurred by the work of Ref.~\cite{Ulf}, suggesting that a peak observed by the LHCb collaboration, which has been accepted as a signal of a pentaquark of hidden charm \cite{LHCb}, was actually due to a triangle singularity. The hypothesis was ruled out in Ref.~\cite{BayarGuo} if the present quantum numbers of this peak hold. Yet, given the fact that some uncertainties concerning these quantum numbers still remain, the issue could be reopened in the future.

The work of Ref.~\cite{BayarGuo} develops a different formalism than the one usually employed, which is very practical and intuitive, and we shall also follow these lines in the present work, which offers a quite different formal derivation of Schmid theorem and allows to see its validity and limitations.

The former recent works on triangle singularities have stimulated many works on the issue \cite{Wang,XieOset,Debastiani,Roca,Sakai,Samart,SakaiRamos,Pavao, SakaiLiang,BayarPavao,LiangSakai,OsetRoca,DaiPavao,GuoZhao,XieGuo, CaoZhao,QinZhao,GuoJuan,YangUlf,LiuUlf}. Further information can be found in the report \cite{GuoReport}.

Along the literature on this topic it is customary to find the statement that due to the Schmid theorem, whenever one has rescattering in the loop to go to the same states as inside the loop, there is no need to evaluate the triangle diagram because its contribution is reabsorbed by the tree level. The purpose of the present paper is to get an insight on the theorem and see where it holds exactly, when it fails and when it is just an approximation and how good or bad can it be. For that, a new derivation is carried out, and a study in terms of the width of the intermediate state $R$ is done.
The failure of the theorem when the $t_{12,12}$ scattering matrix has inelasticities is also shown in detail, providing the quantitative amount of the breaking of the theorem.
Apart from the limit of small $\Gamma_R$, where the theorem strictly holds, we study in a particular case what happens for a realistic width, and a typical $1+2 \to 1+2$ scattering amplitude, which serves as a guide on how much to trust the theorem to eventually neglect the contribution of the triangle loop.

\section{Formulation}

\begin{figure}
  \centering
  \includegraphics[width=0.48\textwidth]{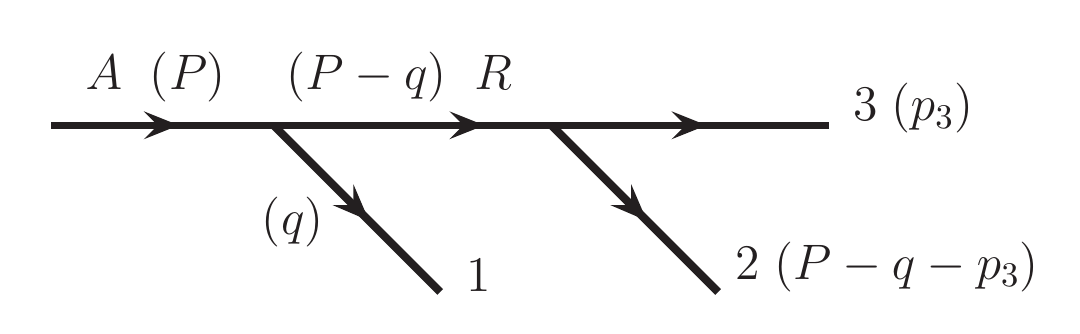}\\
  \caption{Tree level diagram for the process $A \to 1 +2 +3$ mediated by a resonance $R$ that decays into particles $2$ and $3$. In brackets the momenta of the particles.}\label{fig:1}
\end{figure}

Let us study the decay process of a particle $A$ into two particles $1$ and $R$, with a posterior decay of $R$ into particles $2$ and $3$. This is depicted in Fig.~\ref{fig:1}.
We shall also assume for simplicity that all vertices are scalar, and can be represented by just one coupling in each vertex. The conclusions that we will reach are the same for more elaborate couplings, with spin or momentum dependence.

\begin{figure}
  \centering
  \includegraphics[width=0.48\textwidth]{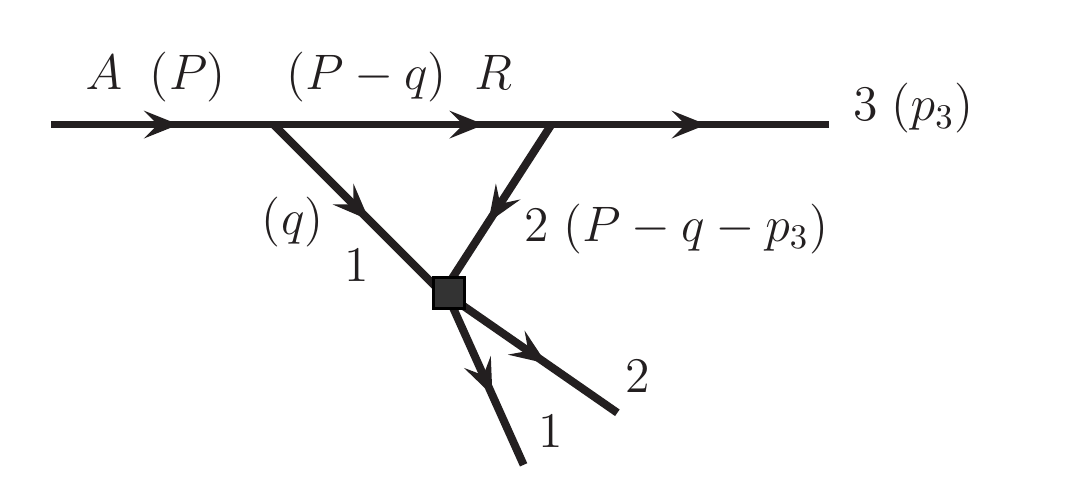}\\
  \caption{Triangle mechanism emerging from the mechanism of Fig.~\ref{fig:1}, with final state interaction of particles $1$ and $2$. In brackets the momenta of the particles.}\label{fig:2}
\end{figure}

The next step is to allow particles $1$ and $2$ to undergo final state interaction, as depicted in Fig.~\ref{fig:2}.
The vertex with particles $1$ and $2$ symbolizes the $1+2 \to 1+2$ scattering matrix. One peculiar property of triangle diagrams is that they can develop a singularity when all the particles inside the loop in Fig.~\ref{fig:2} are placed on shell in the integration and the particles $R$ and $3$ go parallel in the $A$ rest frame \cite{Landau}. The conditions for this to occur, relating the invariant mass of particles $1,2$ with the mass of $A$, can be seen in a pedagogical description of the process in Ref.~\cite{BayarGuo}. It is interesting to note that if particles $R$ and $3$ go in the same direction, then both particles $1$ and $2$ go in the opposite direction to them in the $A$ and $R$ rest frames, respectively.
According to the Coleman-Norton theorem \cite{Coleman} the singularity appears when the classical process of particle $2$, moving in the same direction as particle $1$, but produced later, catches up with particle $1$ and scatters (or fuse to produce another particle).

One should note that in the situation where the $R$ resonance is placed on shell in the triangle loop, as well as particles $1$ and $2$, the tree level mechanism of Fig.~\ref{fig:1} will also have a singularity in the limit of zero width for the resonance $R$, since the amplitude goes as $(M_{\rm inv}(R) -M_R +i\Gamma_R/2)^{-1}$.
However, the tree diagram does not have the restriction that particles $R$ and $3$ should be parallel and hence the region where $R$ can be placed on shell is much wider than for the triangle singularity.

Let us come back to the tree level mechanism of Fig.~\ref{fig:1}. Assume for the moment that the decay into $2+3$ is the only decay channel of the resonance $R$. In the limit of $\Gamma_R \to 0$ one has the decay of $A$ into two elementary particles $1$ and $R$. The width of $A$ can be calculated with the standard formula for decay of two particles or three particles and the results are identical. Intuitively one can say that, once the particle decays into $1$ and $R$, if particle $R$ decays later this does not modify the $A$ width, since this was determined at the moment that $A$ decayed into $1$ and $R$. The same could be said about the triangle mechanism of Fig.~\ref{fig:2}. Once the $A$ particle decays into $1$ and $R$, and $R$ decays into $2$ and $3$, the probability for this process is established and the posterior interaction of $1$ and $2$ should not modify this probability. This argument is the intuitive statement of the Schmid theorem, which technically reads as follow: Let $t_t^{(0)}$ be the $S$-wave projection of the tree level amplitude, $t_t$, of the diagram of Fig.~\ref{fig:1}, evaluated in the rest frame of $1+2$, referred to the angle between the particles $1$ and $3$. Let $t_L$ be the amplitude corresponding to the triangle loop of Fig.~\ref{fig:2}. The Schmid theorem states that
\begin{equation}\label{Schmid}
  t_t^{(0)} +t_L = t_t^{(0)}e^{2i\delta}
\end{equation}
where $\delta$ is the $S$-wave phase shift of the scattering amplitude $1+2 \to 1+2$ (assume also for simplicity that this is the only partial wave in $1+2 \to 1+2$). The formula holds for the case that there is only elastic scattering $1+2 \to 1+2$. In the case that there can be inelastic channels the formula is modified, as we shall see later on, and the Schmid theorem does not strictly hold.

The interesting thing about Eq.~\eqref{Schmid} is that
\begin{equation}
  \left|t_t^{(0)}+t_L\right|^2 = \left|t_t^{(0)}e^{2i\delta}\right|^2 = \left|t_t^{(0)}\right|^2
\end{equation}
and consequently, since $t_t^{(\ell \neq 0)}$ and $t_t^{(0)}$ do not interfere in the angle integration, then
\begin{align}\label{dcos}
& \nonumber   \int_{-1}^{1} d\cos\theta\left|t_t +t_L\right|^2
=
\int_{-1}^{1} d\cos\theta\left|t_t^{(\ell \neq 0)} +t_t^{(0)} +t_L\right|^2 \\
& \nonumber = \int_{-1}^{1} d\cos\theta\left|t_t^{(\ell \neq 0)} +t_t^{(0)}e^{2i\delta}\right|^2   \\
& \nonumber = \int_{-1}^{1} d\cos\theta\left(|t_t^{(\ell \neq 0)}|^2 +|t_t^{(0)}|^2\right) \\
& = \int_{-1}^{1} d\cos\theta|t_t|^2
\end{align}
and as a consequence the contribution of the triangle diagram does not change the width that one obtains just with the tree level diagram.
Note that if we had $1+2 \to 1^\prime+2^\prime$ or $1+2 \to R^\prime$, where $R^\prime$ is some resonance, the theorem does not hold because there is no contribution of the tree level to these reactions.

\subsection{Kinematics of $A \to 1+2+3$}

We assume that all particles are mesons to avoid using different normalization for meson or baryon fields. The width of particle $A$ for this process is given by
\begin{align}\label{width}
 \nonumber  \Gamma_A &= \frac{1}{2M_A}
  \int\frac{d^3p_3}{(2\pi)^3}\frac{1}{2\omega_3}
  \int\frac{d^3p_2}{(2\pi)^3}\frac{1}{2\omega_2}
  \int\frac{d^3p_1}{(2\pi)^3}\frac{1}{2\omega_1}\\
&  \times (2\pi)^4\delta^4(P_A -p_1 -p_2 -p_3)
  \,|t|^2
\end{align}
where $P_A$ is the four-momentum of $A$ and $\omega_i$ the on shell energies $\omega_i = \sqrt{m_i^2 + \vec{p_i}^2}$. We perform the integration over particles $1$ and $2$ in the frame of reference where $\vec{P}_A -\vec{p}_3=0$, where the system of $1+2$ is at rest and the integration over $p_3$ in the $A$ rest frame.
We perform the $d^3p_2$ integration using the $\delta$ function and have in that frame ($\vec{p}_2=-\vec{p}_1$)
\begin{align}
 \nonumber  \Gamma_A &= \frac{1}{2M_A}
  \int\frac{d^3p_3}{(2\pi)^3}\frac{1}{2\omega_3}
  \int\frac{d^3\widetilde{p}_1}{(2\pi)^3}
  \frac{1}{2\widetilde{\omega}_1}
   \frac{1}{2\widetilde{\omega}_2}\\
&  \times 2\pi\delta\left(\widetilde{E}_A -\widetilde{\omega}_3 -\widetilde{\omega}_1(\widetilde{p}_1) -\widetilde{\omega}_2(\widetilde{p}_1)\right)
  \,|t|^2
\end{align}
where the tilde refers to variables in the $1+2$ rest frame, and $t$ is the scattering matrix for the process.

Note that
\begin{equation}
  \widetilde{E}_A -\widetilde{\omega}_3
  = \widetilde{\omega}_1(\widetilde{p}_1) +\widetilde{\omega}_2(\widetilde{p}_1)
  = M_{\rm inv}(12).
\end{equation}
There is no $\phi$ dependence in that frame if we chose $\vec{\widetilde{p}_3}$ in the $z$ direction and the $d^3\widetilde{p}_1$ integration is done with the result
\begin{align}\label{widthResult}
 \Gamma_A = \frac{1}{2M_A}
  \int\frac{d^3p_3}{(2\pi)^3}\frac{1}{2\omega_3}
  \frac{1}{8\pi}
  \int_{-1}^{1}d\cos\theta \, \widetilde{p}_1
  \frac{1}{M_{\rm inv}(12)}
  \,|t|^2
\end{align}
with $\widetilde{p}_1$ the momentum of particle $1$ in the $1+2$ rest frame and $\theta$ the angle between $\vec{\widetilde{p}_1}$ and $\vec{\widetilde{p}_3}$.

Note also that
\begin{equation}\label{Minv23}
  M_{\rm inv}^2(23) = (P_A -p_1)^2 = M_A^2 +m_1^2 -2\widetilde{E}_A\widetilde{E}_1 +2\widetilde{p}_1\widetilde{p}_3\cos\theta.
\end{equation}
Hence, the integration over $\cos\theta$ is equivalent to an integration over $M_{\rm inv}(23)$. Similarly we can write
\begin{equation}\label{Minv12}
  M_{\rm inv}^2(12) = (P_A -p_3)^2 = M_A^2 +m_3^2 -2M_A\omega_3
\end{equation}
where the evaluation has been done in the $A$ rest frame. It is also easy to derive.
\begin{align}
\begin{aligned}
 \widetilde{E}_A & = \frac{1}{2M_{\rm inv}(12)}\left(M_A^2 + M_{\rm inv}^2(12) -m_3^2\right) \\
 \widetilde{E}_3 & = \frac{1}{2M_{\rm inv}(12)}\left(M_A^2 - M_{\rm inv}^2(12) -m_3^2\right) \\
 \widetilde{E}_1 & = \frac{1}{2M_{\rm inv}(12)}\left(M_{\rm inv}^2(12) +m_1^2 -m_2^2\right) \\
 \widetilde{p}_1 & = \frac{\lambda^{1/2}\left(M_{\rm inv}^2(12),m_1^2,m_2^2\right)}{2M_{\rm inv}(12)} \\
 \widetilde{p}_3 & = \frac{\lambda^{1/2}\left(M_A^2,m_3^2,M_{\rm inv}^2(12)\right)}{2M_{\rm inv}(12)}. \\
\end{aligned}
\end{align}
The $d^3p_3$ integration is done in the $A$ rest frame
\begin{equation}
  \int\frac{d^3p_3}{(2\pi)^3}\frac{1}{2\omega_3}
  = \frac{1}{4\pi^2} \int p_3 \, d\omega_3
\end{equation}
and using Eqs.~\eqref{Minv23} and \eqref{Minv12}, the $\omega_3$ and $\cos\theta$ integration can be substituted by integrations over $M_{\rm inv}(12)$ and $M_{\rm inv}(23)$ and one has the formula given in the PDG \cite{PDG}
\begin{align}\label{d2Gam}
 \frac{d^2\Gamma_A}{dM_{\rm inv}(12)dM_{\rm inv}(23)}
  = \frac{1}{64\pi^3}\frac{1}{M_A^3}
  M_{\rm inv}(12)M_{\rm inv}(23)
  \,|t|^2
\end{align}
Yet, for the explanation of the Schmid theorem it is better to use Eq.~\eqref{widthResult}.

\subsection{$t_t$ and $t_L$ amplitudes}

Since we are concerned only with the situation when the particles are close to on shell we will use
\begin{align}\label{prop}
\nonumber  &\frac{1}{q^2 -m^2 +i\epsilon} = \frac{1}{(q^0)^2 -\vec{q\,}^2 -m^2 +i\epsilon}\\
  & = \frac{1}{2\omega(q)}\left\{\frac{1}{q^0 -\omega(q) +i\epsilon} - \frac{1}{q^0 +\omega(q) +i\epsilon}\right\}
\end{align}
and keep only the term $\left[2\omega(q)\left(q^0 -\omega(q) +i\epsilon\right)\right]^{-1}$ since this is the term that can be placed on shell.

The amplitude of Fig.~\ref{fig:1} is given in the $1+2$ rest frame by ($\vec{P}_A = \vec{p}_3$, and we omit the tilde in the momenta)
\begin{equation}\label{tt}
  t_t = g_A \, g_R \, \frac{1}{2\omega_R(\vec{p}_3 -\vec{q}\,)}\frac{1}{\widetilde{E}_A -\omega_1(q) -\omega_R(\vec{p}_3 -\vec{q}\,) +i\epsilon}
\end{equation}
where $g_A, \, g_R$ are the couplings for the decay of $A$ and $R$.
From Eq.~\eqref{tt} we get $t_t^{(0)}$ projecting into $S$-wave
\begin{align}\label{tt0}
 \nonumber  t_t^{(0)}
 &= \frac{1}{2}\int_{-1}^{1}d\cos\theta\frac{1}{2\omega_R(\vec{p}_3 -\vec{q}\,)}\\
  &\times\frac{g_A \, g_R}{\widetilde{E}_A -\omega_1(q) -\omega_R(\vec{p}_3 -\vec{q}\,) +i\epsilon}
\end{align}

Note that in the realistic situation the $i\epsilon$ in Eqs.~\eqref{tt} and \eqref{tt0} will be substituted by $\displaystyle i\frac{\Gamma_R}{2}$.

On the other hand, the amplitude of the triangle diagram can be written in the $1+2$ rest frame as
\begin{align}\label{tloop1}
\nonumber t_L  &= i\int\frac{d^4q}{(2\pi)^4}
  \, \frac{1}{2\omega_R(\vec{p}_3 -\vec{q}\,)}
  \, \frac{1}{\widetilde{E}_A -q^0 -\omega_R(\vec{p}_3 -\vec{q}\,) +i\epsilon}\\
\nonumber  &\times\frac{1}{2\omega_1(q)}
  \, \frac{1}{q^0 -\omega_1(q) +i\epsilon}
  \, \frac{1}{2\omega_2(q)}\\
  &\times  \frac{1}{\widetilde{E}_A -q^0 -\widetilde{E}_3 -\omega_2(q) +i\epsilon}\, g_A \, g_R \, t_{12,12} \, ,
\end{align}
where $t_{12,12}$ is the scattering amplitude for $1+2 \to 1+2$.
The $q^0$ integration is done immediately using Cauchy's theorem and we obtain
\begin{align}\label{tloop2}
\nonumber t_L  &= \int\frac{d^3q}{(2\pi)^3}
  \, \frac{1}{2\omega_R(\vec{p}_3 -\vec{q}\,)}
 \, \frac{1}{\widetilde{E}_A -\omega_1(q) -\omega_R(\vec{p}_3 -\vec{q}\,) +i\epsilon}\\
  &\times \frac{1}{2\omega_1(q)}
  \, \frac{1}{2\omega_2(q)}
 \, \frac{g_A \, g_R \, t_{12,12}}{\widetilde{E}_A -\omega_1(q) -\widetilde{E}_3 -\omega_2(q) +i\epsilon}\, .
\end{align}
We can see that the $R$ propagator term $\left[ 2\omega_R(\vec{p}_3 -\vec{q}\,) \left( \widetilde{E}_A -\omega_1(q) -\omega_R(\vec{p}_3 -\vec{q}\,) +i\epsilon \right)\right]^{-1}$ is common in $t_t$ and $t_L$ of Eqs.~\eqref{tt} and \eqref{tloop2}.
There is, however, a difference since in the loop function $t_L$ the momentum $\vec{q}$ is an integration variable while, in $t_t$, $\vec{q}$ is the momentum of the external particle $1$.
In the loop, after rescattering of particles $1$ and $2$ the momentum $\vec{p}_1$ is different than $\vec{q}$, and only for the situation where all particles in the loop are placed on shell, the moduli of the momenta are equal.

The $\phi$ integral in Eq.~\eqref{tloop2} is trivially done and we have
\begin{align}\label{tloop3}
\nonumber t_L  &= \frac{1}{(2\pi)^2}\int_0^\infty q^2\, dq
 \, \frac{1}{2\omega_1(q)}
 \, \frac{1}{2\omega_2(q)}\\
\nonumber \,&\times \frac{1}{\widetilde{E}_A -\widetilde{E}_3 -\omega_1(q) -\omega_2(q) +i\epsilon}\\
\nonumber &\times \, 2\, \frac{1}{2}
 \int_{-1}^{1}d\cos\theta \, \frac{1}{2\omega_R(\vec{p}_3 -\vec{q}\,)}\\
  &\times\frac{1}{\widetilde{E}_A -\omega_1(q) -\omega_R(\vec{p}_3 -\vec{q}\,) +i\epsilon}
  \, g_A \, g_R \, t_{12,12}\, .
\end{align}
We can see now that the integral over $\cos\theta$ is the same in $t_t^{(0)}$ (Eq.~\eqref{tt0}) and $t_L$ (Eq.~\eqref{tloop3}).

To find the singularity in $t_L$ let us look for the poles of the two propagators. On the one hand from
\begin{equation}
 \widetilde{E}_A -\widetilde{E}_3 -\omega_1(q) -\omega_2(q) +i\epsilon =0
\end{equation}
we obtain
\begin{equation}\label{qon}
  q_{\rm on \, \pm} = \pm \frac{\lambda^{1/2}\left(M_{\rm inv}^2(12),m_1^2,m_2^2\right)}{2M_{\rm inv}(12)} \pm i\epsilon
\end{equation}
On the other hand from
\begin{equation}
\widetilde{E}_A -\omega_1(q) -\omega_R(\vec{p}_3 -\vec{q}\,) +i\epsilon =0
\end{equation}
we get two solutions which depend on $\cos\theta$. Yet we are only interested in $\cos\theta = \pm 1$ because it is there that we will not have cancellations in the principal value of $\displaystyle \int_{-1}^{1}d\cos\theta$ because we cannot go beyond $|\cos\theta\,|=1$ in the integration.
In this case one finds immediately
\footnote{
Note that Eqs.~\eqref{qon} and \eqref{qbar} do not coincide with those in Ref.~\cite{BayarGuo} because 
there
these momenta are obtained in the $A$ rest frame and here in the $1+2$ rest frame. They can be reached by a boost to the frame where $A$ is at rest.
}
\begin{equation}\label{qbar}
  \overline{q} = p_3\cos\theta\frac{E_1^\prime}{M_A}
  \pm \frac{\widetilde{E}_A}{M_A}p_1^\prime \pm i \epsilon,
  \quad (\cos\theta = \pm 1)
\end{equation}
where $E_1^\prime, \, p_1^\prime$ are the energy and momentum of particle $1$  in the $A$ rest frame, while $p_3$ is the momentum of particle $3$ (or $A$) in the $1+2$ rest frame that we have used before. For $\cos\theta = \pm 1$ this is just a boost from the frame where $A$ is at rest ($\overline{q} = p_1^\prime$) to the one where it has a velocity $\displaystyle v_A = \frac{\widetilde{p}_A}{\widetilde{E}_A} = \frac{p_3}{\widetilde{E}_A}$, but the position in the complex plane is given by the $\pm i \epsilon$.

Take now:
\vspace{-7pt}
{\flushleft a) $\cos\theta = -1$}\\

\begin{figure}[h!]
  \centering
  \includegraphics[width=0.35\textwidth]{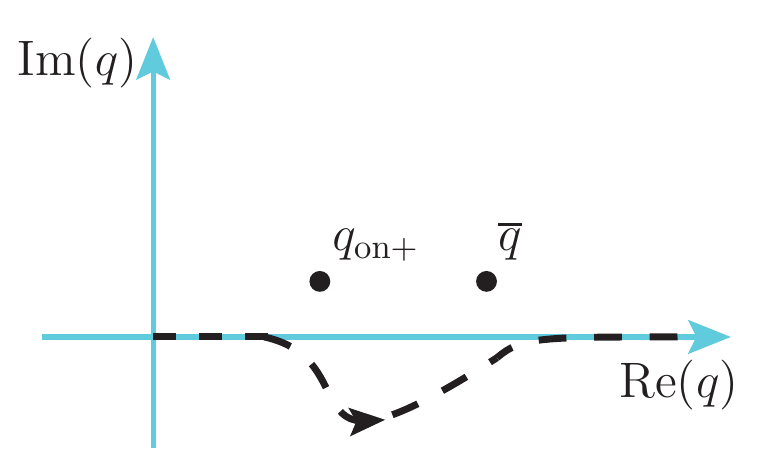}\\
  \caption{Pole positions for the case of $\cos\theta = -1$ when $\overline{q}$ can be positive. The line shows the contour path that one can take to avoid the poles.}\label{fig:3}
\end{figure}

The $(-)$ solution in Eq.~\eqref{qbar},
\begin{equation}\label{Q-}
 Q_- = -p_3\frac{E_1^\prime}{M_A}
  - \frac{\widetilde{E}_A}{M_A}p_1^\prime - i \epsilon,
\end{equation}
gives $\overline{q}$ negative and does not contribute in Eq.~\eqref{tloop3} since $q$ runs from $0$ to $\infty$. If we take the $(+)$ sign,
\begin{equation}\label{Q+}
 Q_+ = -p_3\frac{E_1^\prime}{M_A}
  + \frac{\widetilde{E}_A}{M_A}p_1^\prime + i \epsilon,
\end{equation}
the pole is in the upper side of the complex plane. This situation corresponds to the one in Fig.~\ref{fig:3}. We can see that in that case one can deform the contour path in the $q$ integration to avoid the poles and one does not have a singularity.

{\flushleft b) $\cos\theta = 1$}\\

\begin{figure*}
  \centering
  \includegraphics[width=0.75\textwidth]{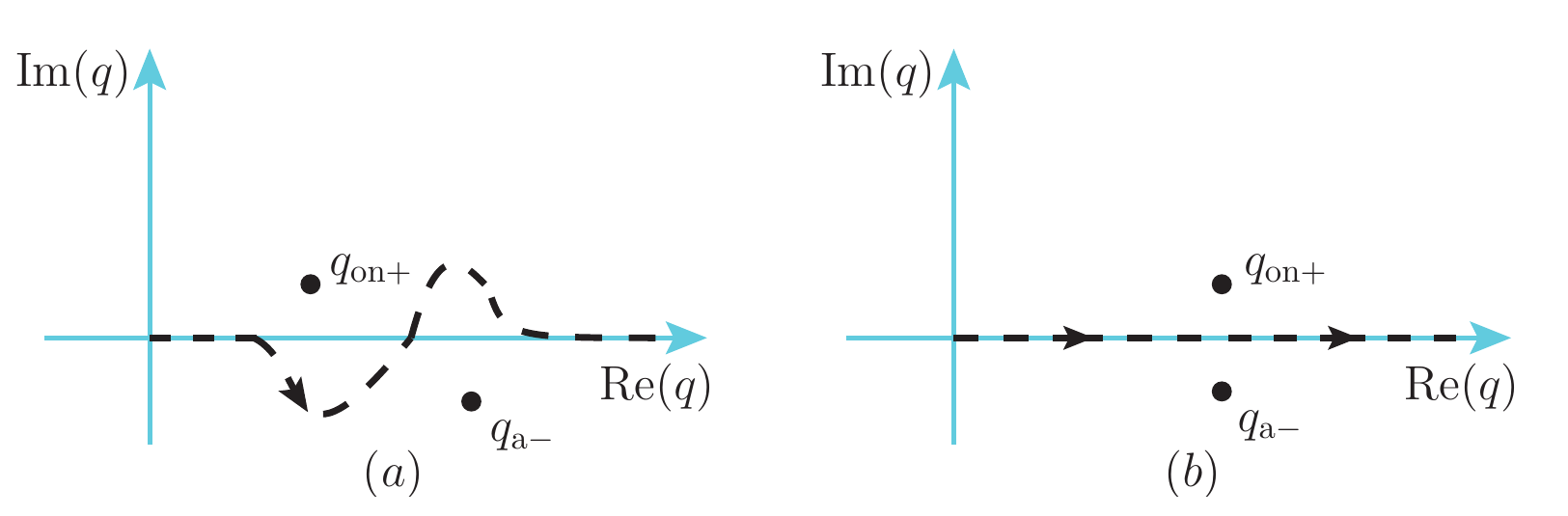}\\
  \caption{Pole positions for $q_{\rm on}$ and $q_{\rm a -}$ for $\cos\theta = 1$. The case (b) leads to the triangle singularity.}\label{fig:4}
\end{figure*}

The $(+)$ solution,
\begin{equation}\label{qaplus}
  q_{\rm a +} = p_3 \frac{E_1^\prime}{M_A}
  + \frac{\widetilde{E}_A}{M_A}p_1^\prime +i \epsilon
\end{equation}
corresponds to a pole in the upper part of the complex plane and does not lead to a singularity. The $(-)$ solution, let us call it $q_{\rm a -}$,
\begin{equation}\label{qaplus}
  q_{\rm a -} = p_3 \frac{E_1^\prime}{M_A}
  - \frac{\widetilde{E}_A}{M_A}p_1^\prime -i \epsilon
\end{equation}
lies in the lower side of the complex plane. In this case if $q_{\rm a -}$ is different from $q_{\rm on}$ of Eq.~\eqref{qon} we have the situation as in Fig.~\ref{fig:4}(a). In this case we can also deform the contour path to avoid the poles in the $q$ integration of $t_L$.
However in the case that
\begin{equation}\label{qonSing}
  q_{\rm on+} = q_{\rm a -} \qquad (\epsilon \to 0),
\end{equation}
corresponding to Fig.~\ref{fig:4}(b), the path of the integral has to go through the poles $q_{\rm on+}$ and $q_{\rm a-}$ and we cannot deform the contour path to avoid the poles. This is the situation of the triangle singularity
\footnote{Another possibility to have singularities is that $q_{\rm a +} = q_{\rm a -} \quad (\epsilon \to 0)$. This leads to a threshold singularity, discussed in Ref.~\cite{BayarGuo}, but which plays no role on the present discussion.}.
Note for further discussions that $Q_- = -q_{\rm a +}$ and  $Q_+ = -q_{\rm a -}$.

\begin{figure}
  \centering
  \includegraphics[width=0.35\textwidth]{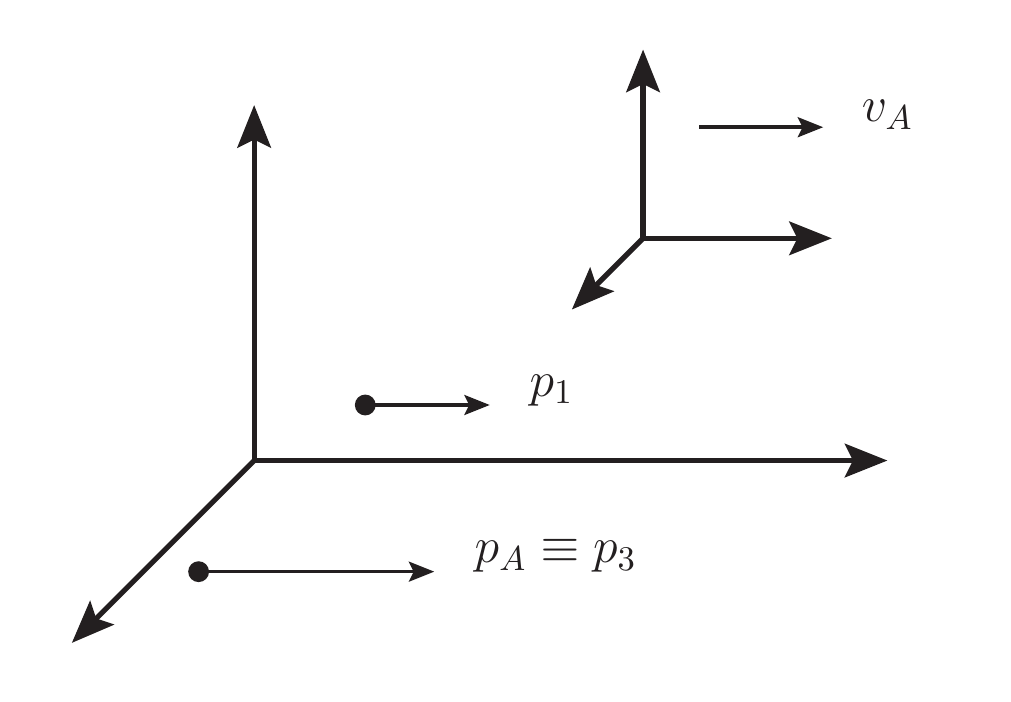}\\
  \caption{Situation of momenta in the $1+2$ rest frame when $\cos\theta = 1$.}\label{fig:5}
\end{figure}

It might be curious to see that we find the singularity for $\cos\theta = 1$, while in Ref.~\cite{BayarGuo} it was found for $\cos\theta = -1$. This is a consequence of the different frame of reference. Indeed, we are in the situation of Fig.~\ref{fig:5}.
We can make a boost with velocity $v_A$ to bring particle $A$ at rest
\begin{equation}
  v_A = \frac{\widetilde{p}_A}{\widetilde{E}_A} = \frac{p_3}{\widetilde{E}_A}\, .
\end{equation}
The velocity of particle $3$ in the $1+2$ rest frame is
\begin{equation}
  v_3 = \frac{p_3}{\widetilde{E}_3} > v_A
\end{equation}
but since the mass of particle $3$ is smaller than $M_A$ then $v_3$ is bigger than $v_A$ and in the boost particle $3$ still goes in the same direction as before.
However $p_1^\prime$ in that frame has a velocity
\begin{equation}
  v_1 = \frac{p_1^\prime}{E_1^\prime}
\end{equation}
and since we are taking the $q_{\rm a-}$ solution positive
\begin{align}
\begin{aligned}
 &  q_{\rm a-} \,=\, p_3 \frac{E_1^\prime}{M_A}
  - \frac{\widetilde{E}_A}{M_A}p_1^\prime
  \,>\, 0 \, ,\\
  & p_3E_1^\prime \,>\, p_1^\prime \widetilde{E}_A \, ;
  \qquad
  \frac{p_3}{\widetilde{E}_A} \,>\, \frac{p_1^\prime}{E_1^\prime}
  \;\Rightarrow\;
  v_A \,>\, v_1 \, .
\end{aligned}
\end{align}
Hence particle $1$ changes direction under the boost of velocity $v_A$ and in the rest frame $\vec{p_3}^\prime$ and $\vec{p_1}^\prime$ have opposite directions, $\cos\theta^\prime = -1$, as it was found in Ref.~\cite{BayarGuo}.
\\

\subsection{Formulation of Schmid theorem}

Let us perform the $\cos\theta$ integration in $t_L$ of Eq.~\eqref{tloop3}.
\begin{align}
\nonumber I(q) & = \int_{-1}^{1} d \cos\theta \,
  \frac{1}{2\omega_R(\vec{p}_3 -\vec{q})} \\
  & \times \frac{1}{\widetilde{E}_A -\omega_1(q) -\omega_R(\vec{p}_3 -\vec{q}) +i\epsilon}.
\end{align}
We can introduce the variable $x$
\begin{align}\label{x}
\nonumber & x  = \omega_R(\vec{p}_3 -\vec{q})
    = \sqrt{m_R^2 + p_3^2 +q^2 -2p_3q\cos\theta} \\
    & d\cos\theta = -\frac{\omega_R}{p_3 q} dx
\end{align}
\begin{align}\label{Iq}
\nonumber I(q) & = -\int_{\omega_R(p_3 +q)}^{\omega_R(p_3 -q)}
    d\,x\,\frac{1}{2\omega_R}\,\frac{\omega_R}{p_3q}\,
    \frac{1}{\widetilde{E}_A -\omega_1(q) -x +i\epsilon}\\
\nonumber & = \int_{\omega_R(p_3 -q)}^{\omega_R(p_3 +q)}\,
     d\,x\,\frac{1}{2p_3q}\,\frac{1}{\widetilde{E}_A -\omega_1(q) -x +i\epsilon}\\
    & = \frac{1}{2p_3q}\ln\left(\frac{\widetilde{E}_A -\omega_1(q) -\omega_R(p_3 -q) +i\epsilon}{\widetilde{E}_A -\omega_1(q) -\omega_R(p_3 +q) +i\epsilon}\right)\, .
\end{align}

It is interesting to see that $I(q)=I(-q)$ is an even function of $q$. Then, since the rest of the integrand of $t_L$ in Eq.~\eqref{tloop3} is also even in $q$ we can write
\begin{align}\label{tloop4}
\nonumber  t_L &= \frac{1}{(2\pi)^2}\frac{1}{2}\int_{-\infty}^{\infty} q^2\, dq\,
  \frac{1}{2\omega_1(q)}\,\frac{1}{2\omega_2(q)}\\
  &\times\frac{g_A \, g_R \, I(q) \, t_{12,12}}{\widetilde{E}_A -\widetilde{E}_3 -\omega_1(q) -\omega_2(q) +i\epsilon}\, .
\end{align}
We should note that the singularity that we are evaluating corresponds to the numerator in Eq.~\eqref{Iq} becoming zero.

\begin{figure}
  \centering
  \includegraphics[width=0.48\textwidth]{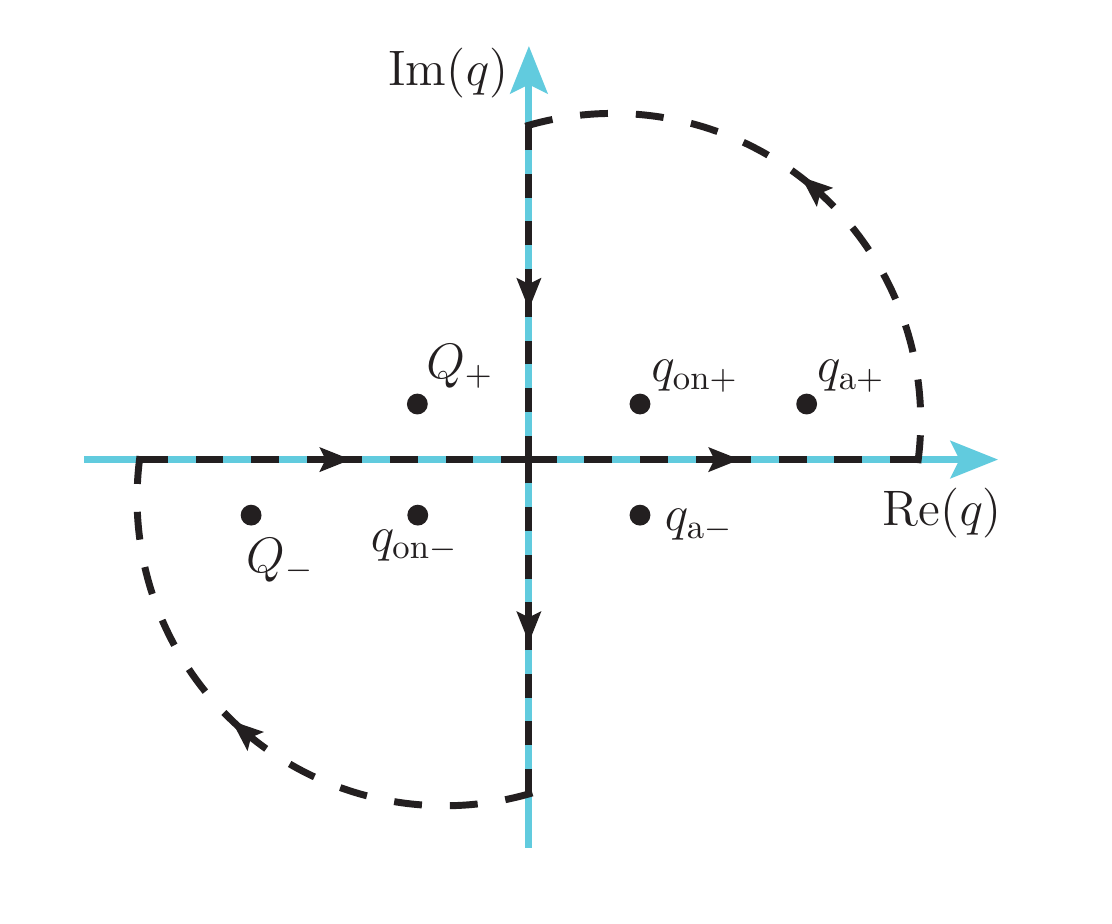}\\
  \caption{Integration path followed to evaluate $t_L$ of Eq.~\eqref{tloop4}.}\label{fig:6}
\end{figure}

Since we have extended the integration to $q$ negative, we must also consider the poles of $Q_-$ and $Q_+$ of Eqs.~\eqref{Q-} and \eqref{Q+}. The situation of the poles is shown in Fig.~\ref{fig:6}.
In this figure we also show the contour path followed to perform the integral over $q$ using Cauchy's theorem. We have
\begin{align}
\nonumber  & \quad
  \int_{-\infty}^{\infty} dq \,\dots
  + \int_{i\infty}^{-i\infty} dq \,\dots
  + \int_{\rm circle} dq \,\dots \\
  &= 2\pi i \,{\rm Res}(q_{\rm on +})
  + 2\pi i \,{\rm Res}(q_{\rm a +})\\
  &- 2\pi i \,{\rm Res}(q_{\rm on -})
  - 2\pi i \,{\rm Res}(Q_-)
\end{align}
The relevant point now is the fact that the integral over the circle of infinity vanishes and $\displaystyle \int_{i\infty}^{-i\infty} dq $,  ${\rm Res}(q_{\rm a +})$ and ${\rm Res}(Q_-)$ do not produce any singularity, and thus
\begin{equation}
   \int_{-\infty}^{\infty} dq \,\dots
   = 2\pi i \,{\rm Res}(q_{\rm on +})
   - 2\pi i \,{\rm Res}(q_{\rm on -})
\end{equation}
is the part of the integral that leads to the singularity.
We go back to Eq.~\eqref{tloop3} and perform the $q$ integration before $\cos\theta$ is integrated.
Since the denominator $\left(\widetilde{E}_A -\widetilde{E}_3 -\omega_1(q) -\omega_2(q)\right)$ vanishes for $q_{\rm on +}$ and $q_{\rm on -}$, we can apply L'H\^{o}pital rule to calculate the residues ($\widetilde{E}_A -\widetilde{E}_3 = M_{\rm inv}(12)$, with $q_{\rm on}$ standing for $q_{\rm on +}$ or $q_{\rm on -}$)
\begin{align}
\nonumber  {\rm Res}(q_{\rm on})
  &\propto \lim_{q\rightarrow q_{\rm on}}  \frac{q -q_{\rm on}}{M_{\rm inv}(12) -\omega_1(q) -\omega_2(q)}\\
  &= \frac{1}{\left(\frac{-q}{\omega_1}\right) +\left(\frac{-q}{\omega_2}\right)}\Bigg{|}_{q_{\rm on }}
  = -\frac{\omega_1\,\omega_2}{q_{\rm on}\,M_{\rm inv}(12)}.
\end{align}

Thus, the singular part of $t_L$  is contained in
\begin{align}
\nonumber  t_L &= \frac{1}{(2\pi)^2}\,\frac{1}{4}\,|q_{\rm on}|\,\frac{1}{M_{\rm inv}(12)}\,
  g_A\, g_R\, t_{12,12}\,(-)4\pi i \\
\nonumber  &\times \frac{1}{2}
 \int_{-1}^{1}d\cos\theta \, \frac{1}{2\omega_R(\vec{p}_3 -\vec{q}\,)}\\
  &\times\frac{1}{\widetilde{E}_A -\omega_1(q) -\omega_R(\vec{p}_3 -\vec{q}\,) +i\epsilon}.
\end{align}
Comparing with Eq.~\eqref{tt0} we find
\begin{equation}\label{tloop5}
  t_L = -i \, \frac{1}{4\pi}\,|q_{\rm on}|\,\frac{1}{M_{\rm inv}(12)}\,t_{12,12}\,t_t^{(0)}.
\end{equation}

We must now recall the relationship of our $t$ matrix in the field theory formulation with the $f$ matrix of Quantum Mechanics
\begin{equation}
f = - \frac{1}{8\pi}\,\frac{1}{M_{\rm inv}}\,t
\end{equation}
which allows us to reformulate Eq.~\eqref{tloop5} as
\begin{equation}
  t_L = 2\, i\, |q_{\rm on}| \, f \, t_t^{(0)}
\end{equation}
and since
\begin{equation}
  f = \frac{\eta \, e^{2i\delta} -1}{2 \, i \, |q_{\rm on}|}
\end{equation}
we find that
\begin{equation}\label{ine}
  t_t^{(0)} +t_L = t_t^{(0)}(1 +2 \, i \, |q_{\rm on}| \, f)
  = \eta\,  e^{2i\delta}\, t_t^{(0)} \, .
\end{equation}
Eq.~\eqref{ine} for the case when there is only the $1,2$ elastic channel ($\eta =1$) is the expression of the Schmid theorem \cite{Schmid}. It was already mentioned in Ref.~\cite{Anisovich} that the Schmid theorem does not hold if $t_{12,12}$ has inelasticities and we can be more quantitative here. Indeed, recalling Eq.~\eqref{dcos} we have now
\begin{align}\label{dcos2}
 \nonumber &\int_{-1}^{1} d\cos\theta\left|t_t +t_L\right|^2
 = \int_{-1}^{1} d\cos\theta
 \left\{
 \left|t_t^{(\ell \ne 0)}\right|^2
 +\eta^2\left|t_t^{(0)}\right|^2
 \right\}\\
 \nonumber&= \int_{-1}^{1} d\cos\theta
 \left\{
 \left|t_t^{(\ell \ne 0)}\right|^2 +\left|t_t^{(0)}\right|^2
 -(1-\eta^2)\left|t_t^{(0)}\right|^2
 \right\}\\
 &= \int_{-1}^{1} d\cos\theta
  \left|t_t\right|^2
 \;
 -(1-\eta^2)
 \int_{-1}^{1} d\cos\theta
 \left|t_t^{(0)}\right|^2
\end{align}
and since $t_t^{(0)}$ contains the same singularity as $t_L$, the singularity of the triangle diagram will show up with a strength of $1-\eta^2$.
\\

\section{Study of the singular behavior}

Let us look at the behavior of $t_L$ around the triangle singularity. For this it is easier to look at $t_t^{(0)}$ which has this same singularity. Let us look at Eq.~\eqref{tt0}. The integral there is $\displaystyle \frac{1}{2} I(q) g_A g_R$, with $I(q)$ from Eq.~\eqref{Iq}. Hence
\begin{equation}\label{ln}
  t_t^{(0)} =\frac{g_A \, g_R}{4 p_3 \, q}\,
  \ln\left(
  \frac{\widetilde{E}_A -\omega_1(q) -\omega_R(p_3 -q) +i\epsilon}{\widetilde{E}_A -\omega_1(q) -\omega_R(p_3 +q) +i\epsilon}
  \right)
\end{equation}
and since for the $q$ of the singularity ($q$ positive from now on)
\begin{equation}
  \widetilde{E}_A -\omega_1(q) -\omega_R(p_3 -q) =0
\end{equation}
then
\begin{equation}
  \widetilde{E}_A -\omega_1(q)
  =\omega_R(p_3 -q)
\end{equation}
and we can write
\begin{align}
\nonumber    t_t^{(0)} &= \frac{g_A \, g_R}{4 p_3 \, q}\,
    \ln \left( \frac{i \, \Gamma_R/2}{\omega_R(p_3 -q) -\omega_R(p_3 +q)}\right)\\
    & \sim \frac{g_A \, g_R}{4 p_3 \, q}\,
    \ln \left(\frac{\Gamma_R}{\omega_R(p_3 -q) -\omega_R(p_3 +q)}\right)
\end{align}
where we have substituted $i\epsilon$ by $i\displaystyle\frac{\Gamma_R}{2}$ and kept the singular part when $\Gamma_R \to 0$. In the $A$ decay width we shall have
\begin{align}\label{GamA}
\nonumber  \frac{d\Gamma_A}{dM_{\rm inv}(12)} &\propto \int_{-1}^{1} d\cos\theta \left|t_t^{(0)}\right|^2\\
  &\sim 2 \left(\frac{g_A \, g_R}{4 p_3 \, q}\right)^2
   \left|\ln \left(\frac{\Gamma_R}{\omega_R(p_3 -q) -\omega_R(p_3 +q)}\right)\right|^2.
\end{align}
On the other hand, we can look at the whole $t_t$ amplitude, and the equivalent part to Eq.~\eqref{GamA} going into the evaluation of the $A$ width is, using Eq.~\eqref{tt},
\begin{align}
 \nonumber &\int_{-1}^{1} d\cos\theta
  \frac{g_A^2 \, g_R^2}{4 \,{\omega_R}^2(\vec{p}_3 -\vec{q}\,)}\\
  &\times \left| \frac{1}{\widetilde{E}_A -\omega_1(q) -\omega_R(\vec{p}_3 -\vec{q}\,) +i\displaystyle \frac{\Gamma_R}{2}} \right|^2.
\end{align}
Performing the same change of variable to the variable $x$ of Eq.~\eqref{x} we obtain
\begin{align}
   \frac{g_A^2 \, g_R^2}{4 p_3 \, q}
   \int_{\omega_R(p_3 -q)}^{\omega_R(p_3 +q)} dx \,\frac{1}{x}\,
   \frac{1}{\left(\widetilde{E}_A -\omega_1(q) -x\right)^2  +\left(\displaystyle\frac{\Gamma_R}{2}\right)^2}.
\end{align}
By making the change $\widetilde{E}_A -\omega_1(q) -x = y$ and substituting $\displaystyle \frac{1}{x}$ by its value in the singularity, $\left[\widetilde{E}_A -\omega_1(q)\right]^{-1}$, we get
\footnote{The integral can also be performed analytically, and the most singular term corresponds to the choice made.}
\begin{align}\label{arctgy}
    -\frac{1}{4 p_3 \, q}\,
   \frac{g_A^2 \, g_R^2}{\widetilde{E}_A -\omega_1(q)}\,
   \frac{2}{\Gamma_R}
    \arctan \frac{2y}{\Gamma_R}\Bigg{|}_{\widetilde{E}_A -\omega_1(q)-\omega_R(p_3 -q)}^{\widetilde{E}_A -\omega_1(q)-\omega_R(p_3 +q)}
\end{align}
where the lower limit in $y$ vanishes. Hence we get
\begin{align}\label{arctg}
\nonumber   & -\frac{1}{4 p_3 \, q}\,
   \frac{g_A^2 \, g_R^2}{\widetilde{E}_A -\omega_1(q)}\,
   \frac{2}{\Gamma_R}\\
    &\times \left\{
    \arctan \frac{2\left(\omega_R(p_3 -q) -\omega_R(p_3 +q)\right)}{\Gamma_R}
     - \arctan \frac{0}{\Gamma_R}
    \right\}
\end{align}
and since $\omega_R(p_3 -q) < \omega_R(p_3 +q)$
\begin{equation}
  \lim_{\Gamma_R \to 0}
  \,\arctan\,
  \frac{2\left(\omega_R(p_3 -q) -\omega_R(p_3 +q)\right)}{\Gamma_R}
  = -\frac{\pi}{2}
\end{equation}
and we get
\begin{equation}\label{GamApi}
  \frac{d\Gamma_A}{dM_{\rm inv}(12)} \propto \frac{1}{4 p_3 \, q}\,
   \frac{g_A^2 \, g_R^2}{\widetilde{E}_A -\omega_1(q)}\,
   \frac{\pi}{\Gamma_R}.\\
\end{equation}
It is interesting to see that when $\Gamma_R \to 0$, where the Schmid theorem holds, $\Gamma_A$ from the tree level, Eq.~\eqref{GamApi}, grows much faster than $\Gamma_A$ from the triangle singularity from Eq.~\eqref{GamA}.
This means, in a certain sense, that the Schmid theorem, even if true, becomes irrelevant, because in the limit where it holds, the contribution to the $A$ width from the tree level is
infinitely much larger than that from the triangle singularity.

Actually, $\Gamma_A$ from Eqs.~\eqref{GamA} and \eqref{GamApi} do not diverge because $g_R$ is related to the width.
In fact, assuming $2+3$ the only decay channel of $R$, for the $S$-wave coupling that we are considering we have
\begin{equation}\label{GamR}
  \Gamma_R = \frac{1}{8\pi}\, \frac{g_R^2}{M_R^2}\, \overline{q}
\end{equation}
with $\overline{q}$ the on shell momentum of particle $2$ or $3$ in $R \to 2+3$ with $R$ at rest.
Then $\Gamma_A$ from Eq.~\eqref{GamApi} goes to a constant, as it should be. If $2+3$ is not the only decay channel then
\begin{equation}
  \frac{g_R^2}{8\pi}\, \frac{1}{M_R^2}\, \overline{q}
  = {\rm BF}(2+3)\Gamma_R
\end{equation}
where ${\rm BF}(2+3)$ is the branching fraction of $R$ to the $2+3$ channel, and the conclusion is the same.
In fact, in the limit $\Gamma_R \to 0$, $\Gamma_A$ calculated with the three body decay is exactly equal to $\Gamma_A$ calculated from $A \to R+1$ times the branching fraction ${\rm BF}(2+3)$.
This means that the contribution of the triangular singularity becomes negligibly small compared to the contribution of the whole tree level.
However, if the width $\Gamma_R$ is different from zero the ratio of contributions of the tree level to the triangle loop becomes finite, and many of the terms that we have been neglecting in the derivation of the Schmid theorem become relevant. Thus, one has to check numerically the contribution of the tree level and the triangle loop, summing them coherently, to see what comes out. This is particularly true when $t_{12,12}$ seats on top of a resonance where the triangle singularity could be very important, or even dominant.

Yet, it is still interesting to note that in the realistic case the reaction studied still has a memory of Schmid theorem, in the sense that the coherent sum of amplitudes gives rise to a width, or cross section, that is even smaller than the incoherent sum of both contributions. One case where one can already check this is in the study of the $\gamma \, p \to \pi^0 \eta \, p$ reaction in Ref.~\cite{Sakai}. The $\gamma \, p$ couples to $\Delta(1700)$ which decays to $\eta \, \Delta(1232)$. The $\Delta$ decays into $\pi^0 p$ and $\eta \, p$ fuse to produce the $N^\ast(1535)$ that subsequently decays into $\eta \, p$.
The contribution of this mechanism is sizable and has been observed experimentally \cite{Gutz}.
In the study of Ref.~\cite{Sakai} one compares the loop contribution with the tree level from $\gamma \, p \to \Delta(1700) \to \eta \, \Delta(1232) \to \eta \pi^0  p$ and they are of the same order of magnitude. A remarkable feature is that the coherent sum of tree level and loop does not change much the contribution of the tree level, even if the loop contribution is sizable, and is smaller than the incoherent sum of both processes (see Fig.~1 of Ref.~\cite{Sakai}). Yet, even then, the $N^\ast(1535)$ contribution to the $\gamma p \to \pi^0 \eta \, p$ gives a distinctive signature, both theoretically and experimentally. The message is clear: one must evaluate both the tree level and the loop contribution for each individual case to assert the relevance of the triangle singularity. Obviously in the case that the final channel is different from the internal one of the loop, there is no contribution of tree level and then the triangle singularity shows up clearly.
\\

\section{Considerations on the Dalitz plot}

In Fig.~\ref{fig:7} we show the Dalitz plot for a typical $A \to 1+2+3$ process, where $1,2,3$ are final states on shell. 
Recalling Eq.~\eqref{Minv23} that we reproduce again here
\begin{equation}\label{Minv23again}
  M_{\rm inv}^2(23) = M_A^2 +m_1^2 -2\widetilde{E}_A\widetilde{E}_1 +2\widetilde{p}_1\widetilde{p}_3\cos\theta
\end{equation}
and that $\widetilde{E}_A,\widetilde{E}_1,\widetilde{p}_1,\widetilde{p}_3$ all depend on $M_{\rm inv}(12)$, by fixing $M_{\rm inv}(12)$ Eq.~\eqref{Minv23again} gives the limits of the Dalitz plot. For $\cos\theta= -1$ we have the lower limit and for $\cos\theta= 1$ we have the upper limit of the boundary of the Dalitz region. Let us cut the Dalitz boundary with a line of $M_{\rm inv}(23) = M_R$. This line cuts the boundary in points $A$ and $D$.
In these points
we have all final particles on shell and $M_{\rm inv}(23) = M_R$. This is the situation of a triangle singularity provided that Eq.~\eqref{qonSing} is fulfilled (note that there are two solutions of $q$ for $\cos\theta=1$, but only one produces the triangle singularity, where in addition Eq.~\eqref{qonSing} is fulfilled). On the other hand, from point $A$ to $D$, $M_{\rm inv}(23) = M_R$ is allowed for some valid $\cos\theta$ and we should expect a singularity (up to the factor  $g_R^2$) of the tree level mechanism. Let us study this in detail.

In Eq.~\eqref{GamApi} we already saw the contribution of $t_t$ to the $A$ width evaluating $\displaystyle \int_{-1}^{1}d\cos\theta\left|t_t\right|^2$ at the point of the triangle singularity (point $D$ in Fig.~\ref{fig:7}).
Let us now evaluate it for an invariant mass $M_{\rm inv}(12)$ between points $A$ and $D$ in the figure. Close to point $A$ to the left we shall have now
\begin{equation}\label{2y}
  \int_{-1}^{1}d\cos\theta\left|t_t\right|^2
  = -\frac{g_A^2 \, g_R^2}{4 p_3 \, q}\,
   \frac{1}{\widetilde{E}_A -\omega_1(q)}\,
   \frac{2}{\Gamma_R}
    \arctan \frac{2y}{\Gamma_R}\Bigg{|}_{y_2}^{y_1}
\end{equation}
with $y_1<0$ and $y_2>0$, and hence
\begin{align}
\nonumber  \lim_{\Gamma_R \to 0} \,\int_{-1}^{1}d\cos\theta\left|t_t\right|^2
  &= -\frac{g_A^2 \, g_R^2}{4 p_3 \, q}\,
   \frac{1}{\widetilde{E}_A -\omega_1(q)}\,
   \frac{2}{\Gamma_R}\\
   &\times \left( -\frac{\pi}{2} -\frac{\pi}{2}\right)
\end{align}
which is twice the result of Eq.~\eqref{GamApi}.
This value stabilizes as we move from $A$ to $D$, and the interesting thing is that it goes like $(\Gamma_R)^{-1}$ (apart from the factor $g_R^2$).
In addition we can evaluate $\displaystyle\int_{-1}^{1}d\cos\theta \left|t_t\right|^2$ for $M_{\rm inv}(12)$ bigger than the one corresponding to point $A$. In this case $R$ is never on shell and in the limit of small $\Gamma_R$ we get zero contribution relative to the on shell one. Technically one would find it from Eq.~\eqref{2y} since now both $y_1$ and $y_2$ would be negative and the upper and lower limits of $\displaystyle\arctan \frac{2y}{\Gamma_R}$ cancel.
\\

\section{Results}

We are going to perform calculations with a particular case with the following configuration:
\begin{align}\label{Ms}
\begin{aligned}
  M_A & = 2154 \, {\rm MeV}\\
  M_R & = 1600 \, {\rm MeV},
  \quad   \Gamma_R  = 30 \, {\rm MeV}\\
  M_1 & = 500 \, {\rm MeV}\\
  M_2 & = 200 \, {\rm MeV}\\
  M_3 & = 900 \, {\rm MeV}.
\end{aligned}
\end{align}
In addition we shall take an amplitude $1+2 \to 1+2$ parameterized in terms of a Breit-Wigner form
\begin{equation}\label{BW}
  t_{12,12} = \frac{g^2}{M_{\rm inv}^2(12) -M_{\rm BW}^2 +i \, \Gamma_{BW}(M_{\rm inv}(12))M_{\rm inv}(12)}
\end{equation}
with $M_{\rm BW} = 800$ MeV, $\Gamma_{BW} = 20$ MeV.
We shall use Eq.~\eqref{qon} and choose $g^2$ such that the width is given by Eq.~\eqref{GamR}.
In this case we have purely elastic scattering and $g=1338.7$ MeV. We will also consider the case where we take the same value of $g$ and a width double the elastic one to account for inelasticities, to test what happens in the case of inelastic channels.

We will use Eqs.~\eqref{tt} and \eqref{tloop3}  and integrate Eq.~\eqref{d2Gam} over $M_{\rm inv}(23)$ to obtain $d\Gamma_A/d M_{\rm inv}(12)$.
Note that according to Eq.~\eqref{Minv23} the integral over $\cos\theta$ that we have done is simply $2 M_{\rm inv}(23) d M_{\rm inv}(23) / 2 \widetilde{p}_1 \widetilde{p}_3$.
The limits of integration are obtained from Eq.~\eqref{Minv23} for $\cos\theta = \pm 1$, and explicit formulas can be obtained from the PDG \cite{PDG}. In Eq.~\eqref{tloop3} we have used a cutoff in the $q$ integration of $600$ MeV, a common value in many of these problems.

The choice of variables in Eqs.~\eqref{Ms} and \eqref{BW} is done such that Eq.~\eqref{qonSing} is satisfied and we have a triangle singularity for this configuration.

\begin{figure}
  \centering
  \includegraphics[width=0.48\textwidth]{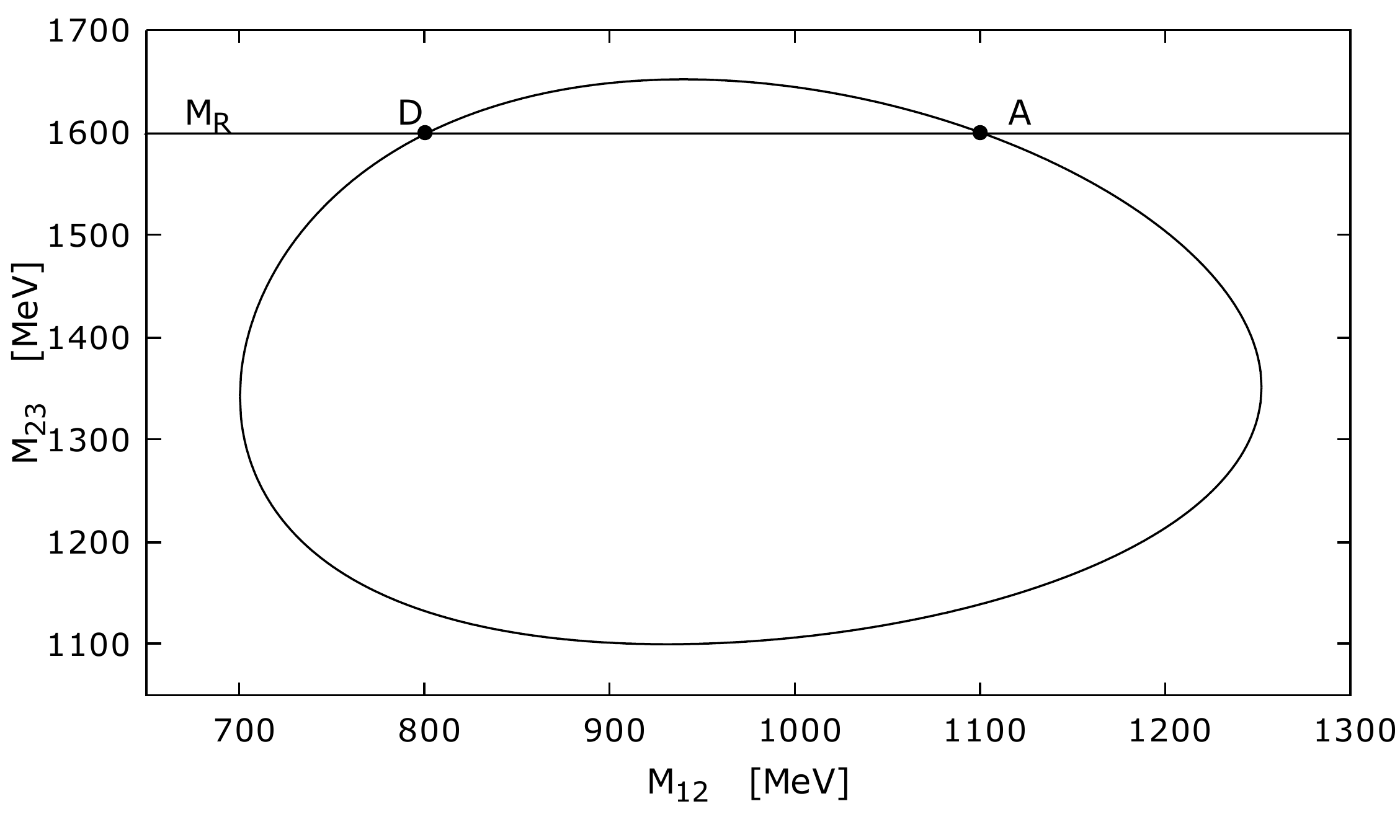}\\
  \caption{Dalitz plot for $A \to 1+2+3$ with the parameters of Eqs.~\eqref{Ms} and \eqref{BW}. Point $D$ corresponds to the triangle singularity.}\label{fig:7}
\end{figure}

In Fig.~\ref{fig:7} we show the Dalitz plot for the reaction $A \to 1+2+3$. We can indeed see that the triangle singularity point corresponds to point $D$ of Fig.~\ref{fig:7}.

In the first place we evaluate $d\Gamma_A/d M_{\rm inv}(12)$ for $M_A =2200$ MeV, about $50$ MeV above the mass that leads to a triangle singularity at $M_{\rm inv}(12) = 800$ MeV.
With this value of $M_A$, the triangle singularity appears at $M_{\rm inv}(12) = 769$ MeV, about $30$ MeV below the Breit-Wigner mass.
We can expect that the loop function will give a small contribution, since $t_{12,12}$ at this invariant mass is significantly reduced compared to its peak. Yet, this serves us to investigate the points discussed in the former section.

\begin{figure}
  \centering
  \includegraphics[width=0.48\textwidth]{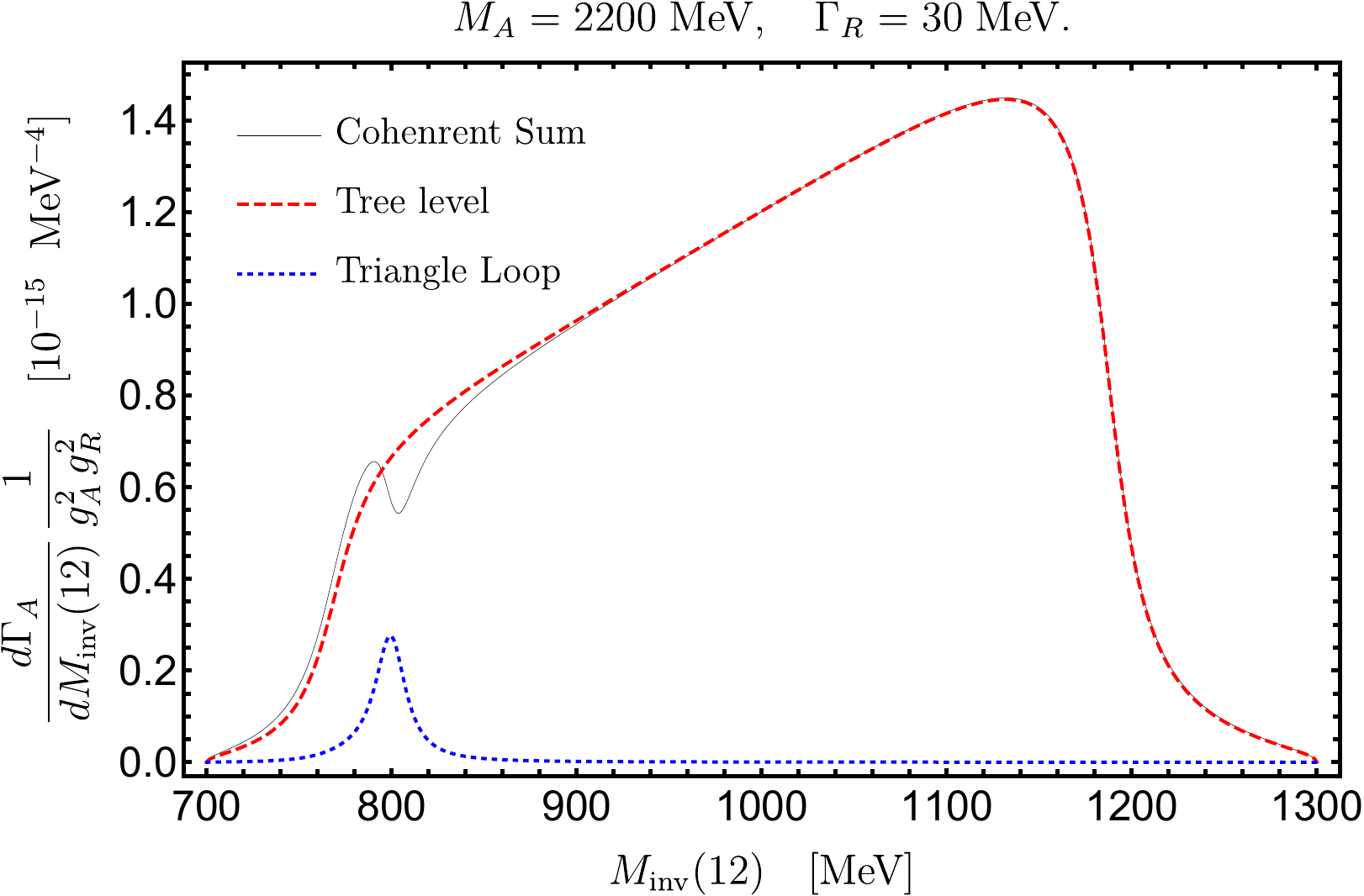}\\
  \caption{$\displaystyle \frac{d\Gamma_A}{d M_{\rm inv}(12)}\,\frac{1}{g_A^2 \, g_R^2}$ as a function of $M_{\rm inv}(12)$ with $M_A = 2200$ MeV and $\Gamma_R = 30$ MeV.}\label{fig:8}
\end{figure}

In Fig.~\ref{fig:8} we show $d\Gamma_A/d M_{\rm inv}(12)/g_A^2 \, g_R^2$ as a function of $M_{\rm inv}(12)$. We show the results for the triangle diagram alone, the tree level alone, and the coherent sum for $\Gamma_R = 30$ MeV.
As we have discussed previously, we have proved the Schmid theorem in the limit of $\Gamma_R \to 0$. Fig.~\ref{fig:8} gives us the answer of what happens when we look at a realist case with $\Gamma_R$ of the order of tens of MeV. As we can see, the triangle loop gives a sizable contribution of the order of the tree level, and the coherent sum of the triangle diagram and the tree level diagram gives rise to very distinct structure, as a consequence of the resonance in the $1,2$ channel enhanced by the triangle diagram. This already tell us that in a realistic calculation we should not rely on the Schmid theorem to neglect the triangle diagram with elastic rescattering of the internal particles of the loop.

\begin{figure}
  \centering
  \includegraphics[width=0.48\textwidth]{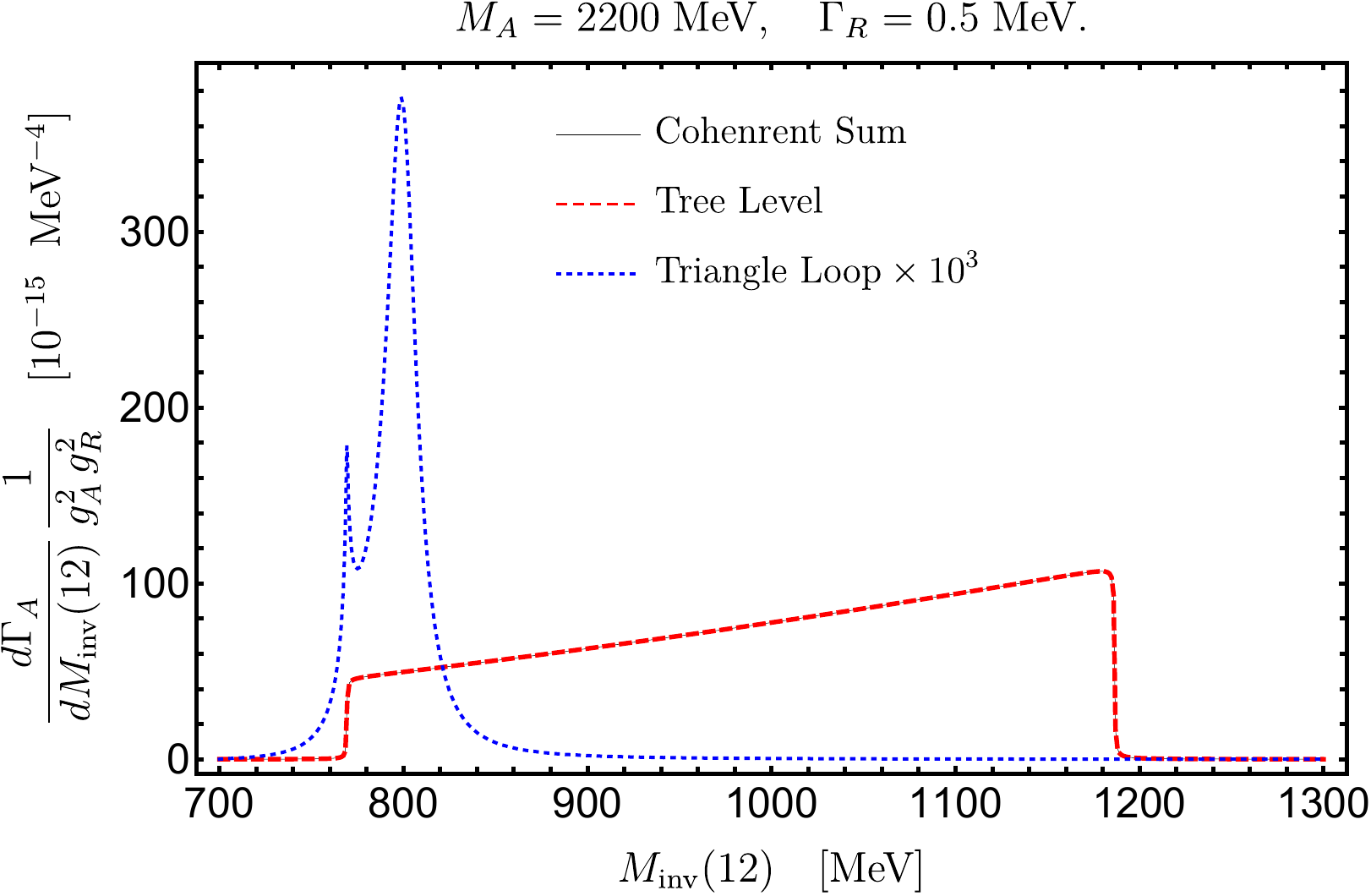}\\
  \caption{$\displaystyle \frac{d\Gamma_A}{d M_{\rm inv}(12)}\,\frac{1}{g_A^2 \, g_R^2}$ as a function of $M_{\rm inv}(12)$ with $M_A = 2200$ MeV and $\Gamma_R = 0.5$ MeV. The dashed and solid lines are indistinguishable in the figure.}\label{fig:9}
\end{figure}

\begin{figure}
  \centering
  \includegraphics[width=0.48\textwidth]{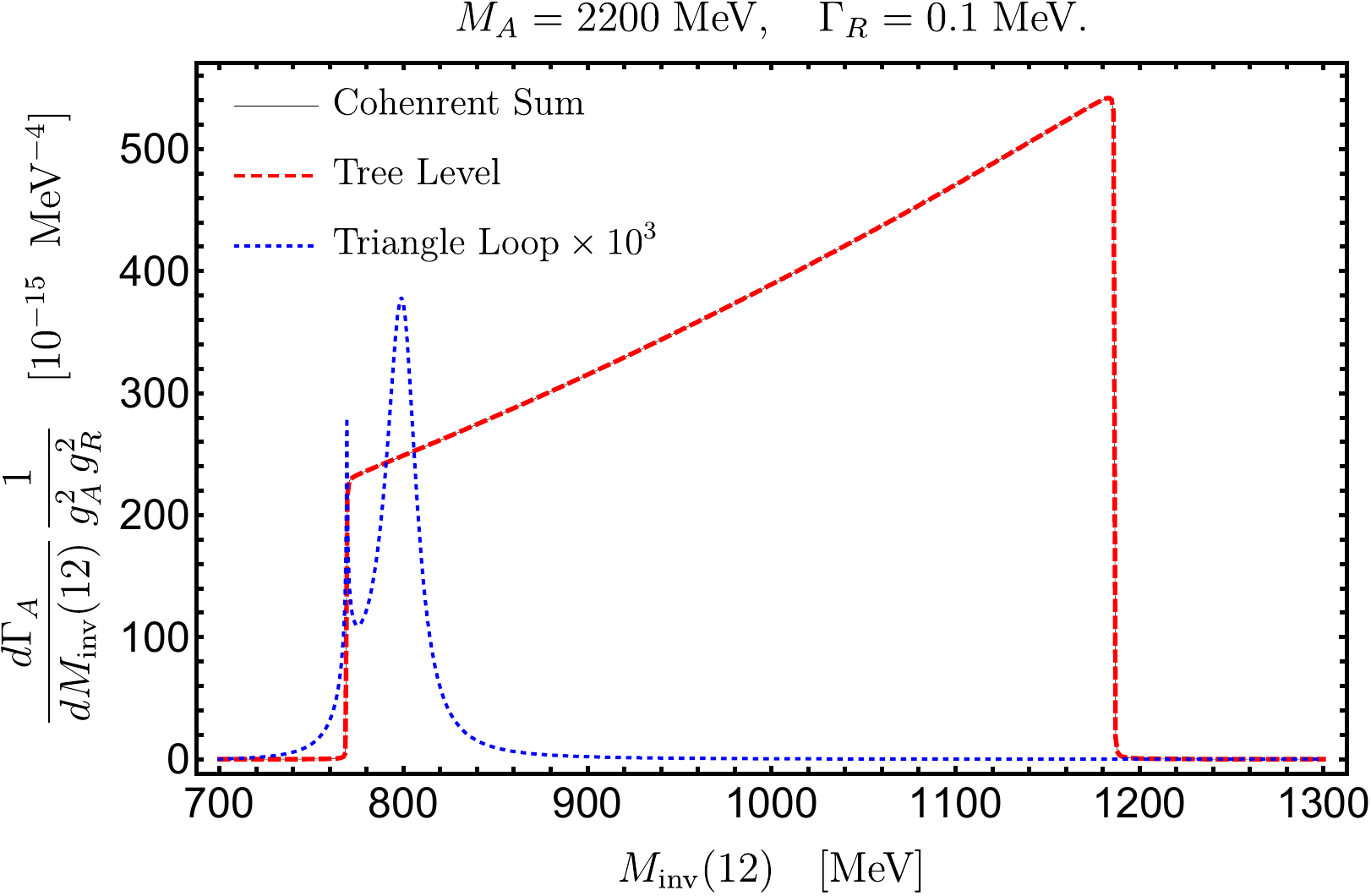}\\
  \caption{$\displaystyle \frac{d\Gamma_A}{d M_{\rm inv}(12)}\,\frac{1}{g_A^2 \, g_R^2}$ as a function of $M_{\rm inv}(12)$ with $M_A = 2200$ MeV and $\Gamma_R = 0.1$ MeV. The dashed and solid lines are indistinguishable in the figure.}\label{fig:10}
\end{figure}

In Fig.~\ref{fig:9} we show the same results but calculated with $\Gamma_R = 0.5$ MeV.
As we can see, the triangle singularity gives a small contribution compared to the tree level, and the coherent sum of the two does not change the contribution of the tree level, as expected from the Schmid theorem.
In Fig.~\ref{fig:10} we show the same results but calculated with $\Gamma_R = 0.1$ MeV.
Comparing these results with those of Fig.~\ref{fig:9}, we can see that as $\Gamma_R$ is made smaller the tree level contribution grows more or less like $1/\Gamma_R$, as expected from Eq.~\eqref{GamApi}, while the triangle contribution grows much less.

\begin{figure}
  \centering
  \includegraphics[width=0.48\textwidth]{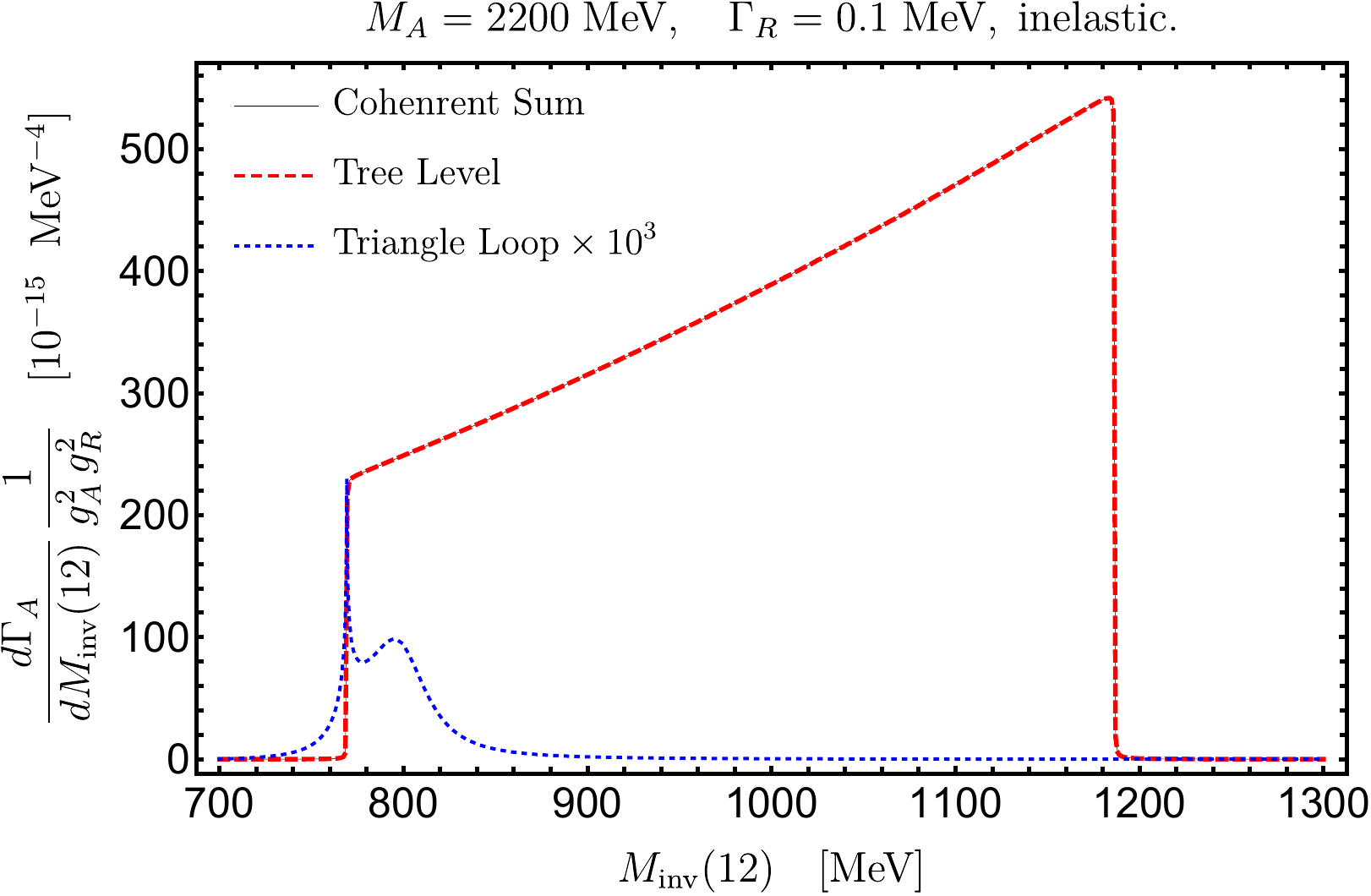}\\
  \caption{$\displaystyle \frac{d\Gamma_A}{d M_{\rm inv}(12)}\,\frac{1}{g_A^2 \, g_R^2}$ as a function of $M_{\rm inv}(12)$ with $M_A = 2200$ MeV and $\Gamma_R = 0.1$ MeV. Here $\Gamma_{BW} \to 2\Gamma_{BW}$ to account for the inelasticities. The dashed and solid lines are indistinguishable in the figure.}\label{fig:11}
\end{figure}

In Fig.~\ref{fig:11} we show again the results for the same $M_A$ mass and $\Gamma_R = 0.1$ MeV, but this time we take in Eq.~\eqref{BW} $\Gamma_{BW} \to 2\Gamma_{BW}$ to account for the inelasticities.
The contribution of the triangle loop is reduced with respect to the one in the elastic case of Fig.~\ref{fig:10}. Curiously, the relative strength of the singularity (narrow peak in the dotted line) with respect to the resonance peak increases now, as a reminder that according to Eq.~\eqref{dcos2} the Schmid theorem does not hold. Yet, the main message from Figs.~\ref{fig:9}, \ref{fig:10} and \ref{fig:11} is that in the limit of $\Gamma_R \to 0$, where the Schmid theorem holds, the relative strength of the triangle diagram versus the tree level becomes negligible.

\begin{figure}
  \centering
  \includegraphics[width=0.48\textwidth]{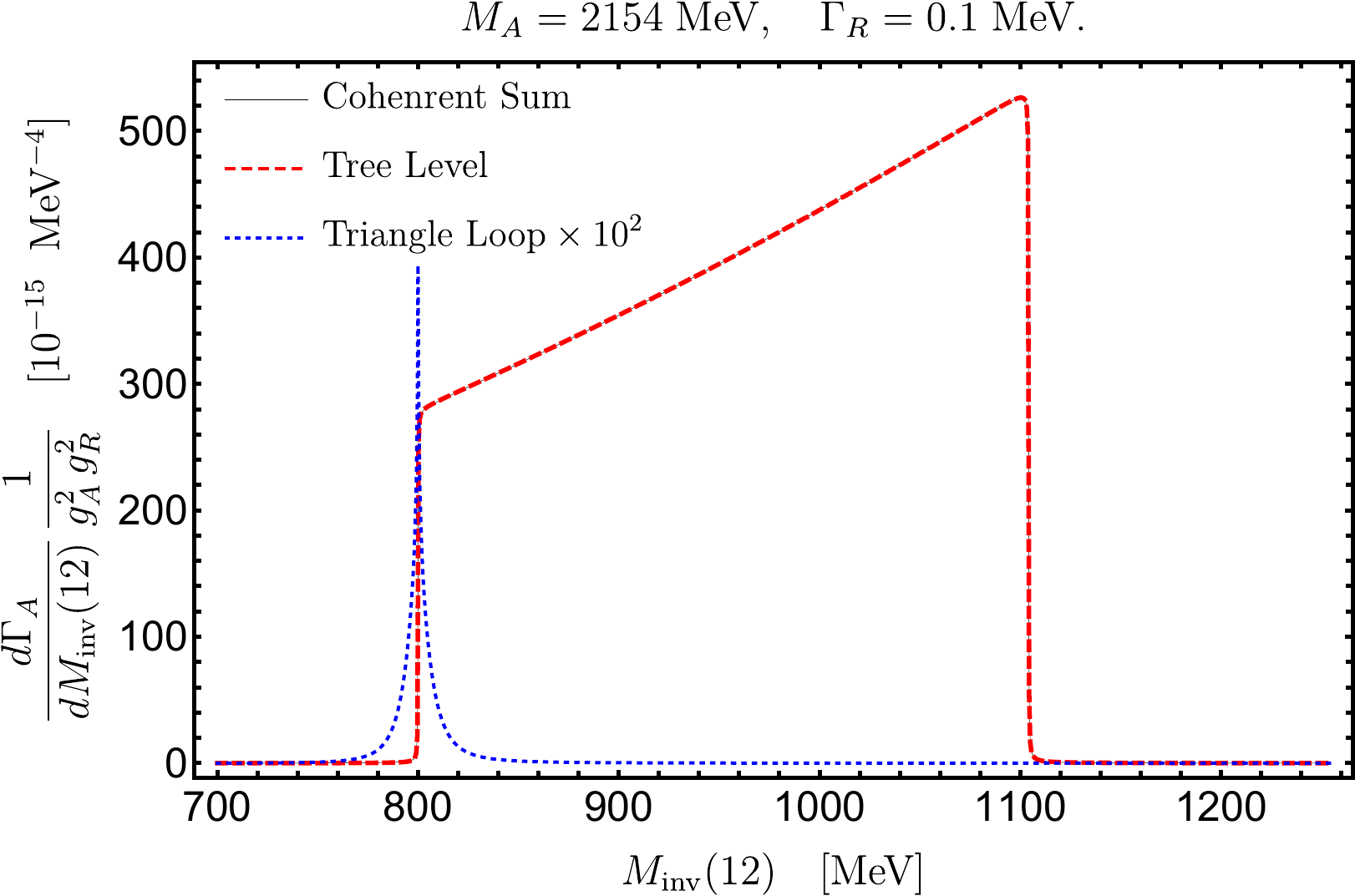}\\
  \caption{$\displaystyle \frac{d\Gamma_A}{d M_{\rm inv}(12)}\,\frac{1}{g_A^2 \, g_R^2}$ as a function of $M_{\rm inv}(12)$ with $M_A = 2154$ MeV and $\Gamma_R = 0.1$ MeV. The dashed and solid lines are indistinguishable in the figure.}\label{fig:12}
\end{figure}

Next we perform the calculations for $M_A = 2154$ MeV, first for $\Gamma_R=0.1$ MeV, and show the results in Fig.~\ref{fig:12}.
The novelty in this case is that, since the triangle singularity occurs for $M_{\rm inv}(12) =800$ MeV, equal to $M_{\rm BW}$, now the contribution of the triangle mechanism is much bigger than in the former cases.
Yet, relative to the tree level its strength is very small and in the coherent sum one does not appreciate its contribution, as was the case of Figs.~\ref{fig:9} and \ref{fig:10}, in agreement with the Schmid theorem.
Actually this is a good case to show the effects of the Schmid theorem, since the strength of the peak is about $10^{-2}$ the one of the tree level. This implies a factor $10^{-1}$ in the amplitude, and in an ordinary coherent sum of these two amplitudes one might expect a contribution of about $20$\% in the differential width, which is not the case in Fig.~\ref{fig:12}.
Yet we should stress once more that the triangle mechanism becomes negligible on its own, independently of the Schmid theorem, when $\Gamma_R \to 0$.

\begin{figure}
  \centering
  \includegraphics[width=0.48\textwidth]{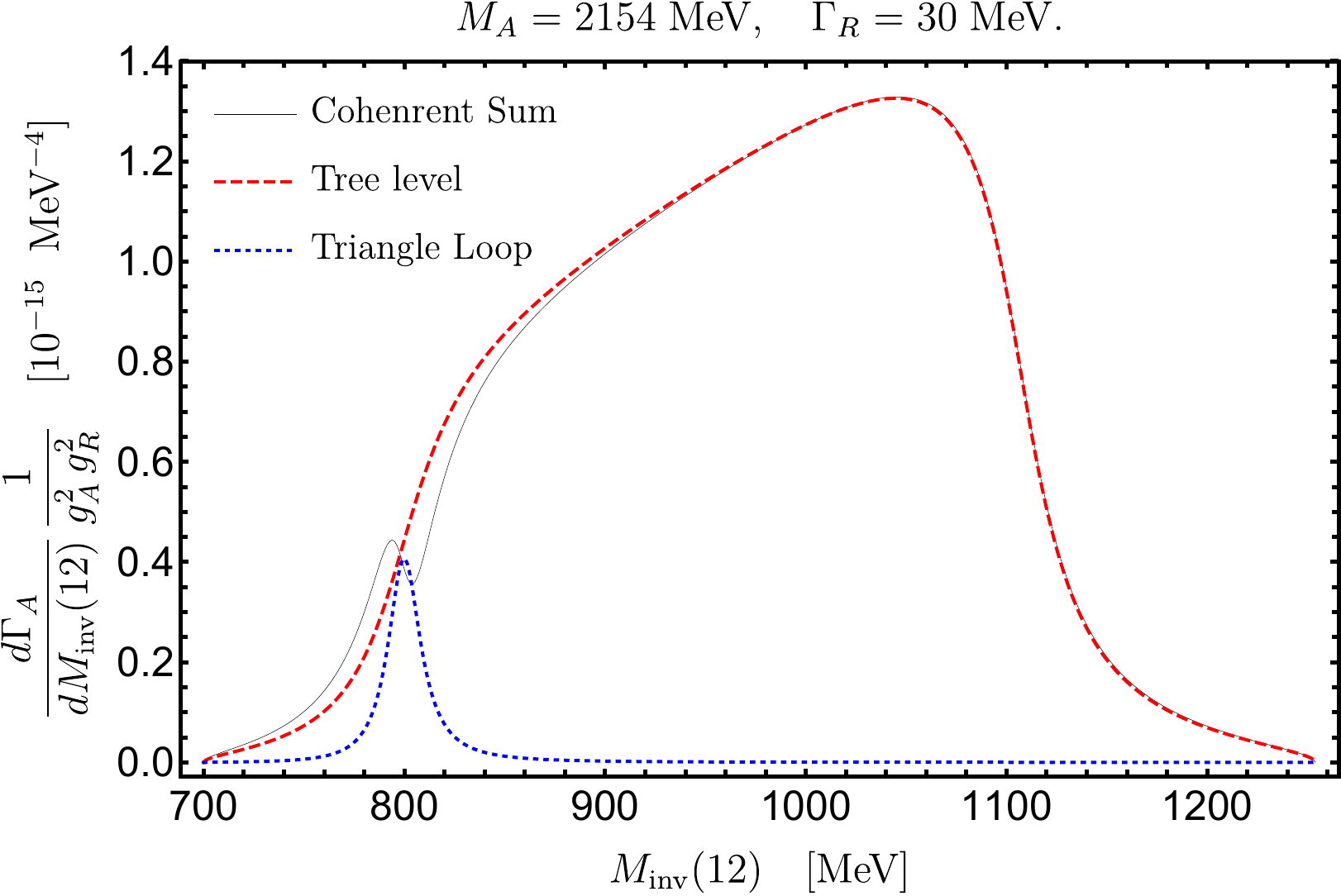}\\
  \caption{$\displaystyle \frac{d\Gamma_A}{d M_{\rm inv}(12)}\,\frac{1}{g_A^2 \, g_R^2}$ as a function of $M_{\rm inv}(12)$ with $M_A = 2154$ MeV and $\Gamma_R = 30$ MeV.
  }\label{fig:13}
\end{figure}

In Fig.~\ref{fig:13} we show the same results now for $\Gamma_R =30$ MeV. We can see now that the effect of the triangle singularity shows in $\displaystyle \frac{d\Gamma_A}{d M_{\rm inv}(12)}$. However, it is interesting to see that the coherent sum of the tree level and the triangle singularity is smaller than their incoherent sum, indicating that, although one is in a region where the Schmid theorem does not strictly hold, the process still has some memory of the the absorption of the triangle mechanism by the tree level amplitude that occurs in the limit of small $\Gamma_R$.
We should also note that if we make $\Gamma_R$ bigger the relative strength of the loop contribution to the tree level grows and can become dominant at the invariant mass of the triangle singularity.

\section{Conclusions}

We have done a new derivation of the Schmid theorem and have studied the results as a function of $\Gamma_R$, the width of the intermediate state in the triangle loop that decays into an external particle and an internal one. The Schmid theorem holds strictly in the limit when $\Gamma_R \to 0$. We show this again and illustrate it with a numerical example.

The first thing that we find is that when $\Gamma_R \to 0$ the relative weight of the triangle singularity versus the tree level contribution goes to zero. This means that, in a strict sense, the Schmid theorem would be irrelevant because where it holds and shows that the singularity changes only the phase of the $S$-wave part of the tree level amplitude, $t_t^{(0)}$, the strength of the triangle diagram or of $t_t^{(0)}$ is very small compared with the whole contribution of the full tree level amplitude.

We conducted some tests to see what happens when we have $\Gamma_R$ finite and also when the scattering amplitude of the two particles that interact in the loop sit on top of a resonance.

In general lines we see that for finite $\Gamma_R$ widths the Schmid theorem does not strictly hold but the $A \to 1+2+3$ process still has some memory of the theorem in the sense that the coherent sum of tree level and triangle singularity gives rise to a differential width where the interference is much smaller than usual, with final results even smaller than the incoherent sum of the two mechanisms.

We also saw that in the case of a $1+2 \to 1+2$ scattering amplitude with inelasticities the Schmid theorem does not hold and we could quantize the contribution of the triangle singularity, both analytically and numerically in the example we discussed.

The biggest contribution of the triangle singularity appears when the $1+2 \to 1+2$ scattering amplitude seats on top of a resonance. In this case, and for finite $\Gamma_R$, the contribution of the triangle singularity shows up clearly and can be even dominant over the tree level in some cases.

The message in general is that for each particular case one has to calculate both the tree level and the triangle mechanism and sum them coherently to see what comes out. Invoking the Schmid theorem to rule out the triangle mechanism in the case that the rescattering of particles $1,2$ goes to the same channel should not be done.

\begin{acknowledgments}
V. R. Debastiani acknowledges the Programa Santiago Grisolia of Generalitat Valenciana (Exp. GRISOLIA/2015/005). S.~Sakai acknowledges the support by NSFC and DFG through funds provided to the Sino-German CRC110 ``Symmetries and the Emergence of Structure in QCD'' (NSFC Grant No. 11621131001), by the NSFC (Grant No. 11747601), by the CAS Key Research Program of Frontier Sciences (Grant No. QYZDB-SSW-SYS013) and by the CAS Key Research Program (Grant No. XDPB09).
This work is also partly supported by the Spanish Ministerio de Economia y Competitividad and European FEDER funds under the contract number FIS2014-57026-REDT, FIS2014-51948-C2-1-P, and FIS2014-51948-C2-2-P, and the Generalitat Valenciana in the program Prometeo II-2014/068.
\end{acknowledgments}


  \end{document}